\documentclass[11pt]{article}
\usepackage{fullpage}

\usepackage{amssymb}
\usepackage{graphicx}
\usepackage{url,amssymb}

\def\arccosh{\mathrm{arccosh}}

\def\cosh{\mathrm{cosh}}

\def\calP{\mathcal{P}}
\def\calB{\mathcal{B}}

\def\calU{\mathcal{U}}
\def\calK{\mathcal{K}}
\def\calL{\mathcal{L}}
\def\LERP{\mathrm{LERP}}

\def\Bi{\mathrm{Bi}}
\def\vor{\mathrm{Vor}}
\def\bbX{\mathbb{X}}
\def\bbH{\mathbb{H}}
\def\bbR{\mathbb{R}}
\def\bbB{\mathbb{B}}

\def\bbL{\mathbb{L}}
\def\bbC{\mathbb{C}}

\def\DT{\mathrm{DT}}

\def\inner#1#2{{\langle #1,#2\rangle}}

\def\InnerE#1#2{{\left\langle #1,#2\right\rangle}_\mathrm{E}}
\def\innerE#1#2{{\langle #1,#2\rangle}_\mathrm{E}}
\def\innerL#1#2{{\langle #1,#2\rangle}_\mathrm{L}}

\def\ceil#1{{\lceil {#1}\rceil}}

\def\ballE{\mathrm{Ball}_E}

\def\ds{\mathrm{d}s}
\def\dx{\mathrm{d}x}

\def\PSL{\mathrm{PSL}}

\def\pow{\mathrm{\Pi}}
\def\path#1{}
\def\calS{\mathcal{S}}

\newtheorem{theorem}{Theorem}

\begin{document}

\title{The hyperbolic Voronoi diagram in arbitrary dimension\footnote{In this revised manuscript, we added the geodesic equation for the Klein geodesics and the connection of hyperbolic geometry with the special spaces of symmetric positive-definite matrices with prescribed determinants.}}

\author{Frank Nielsen\thanks{Corresponding author. Fax:(+81) 3-5448-4380.
\protect\url{Frank.Nielsen@acm.org}.
Sony Computer Science Laboratories Inc., 3-14-13 Higashi Gotanda 3F, Shinagawa-Ku, 
Tokyo 141-0022, Japan.}
\and
Richard Nock\thanks{\protect\url{Richard.Nock@martinique.univ-ag.fr}. CEREGMIA UAG,
Campus de Schoelcher, BP 7209. 97275 Schoelcher, 
Martinique, France.}
}

\date{}

\maketitle

\begin{abstract}
We show that in the Klein ball model of hyperbolic space, the hyperbolic Voronoi diagram is affine and amounts to clip a corresponding power diagram, requiring however algebraic arithmetic.
By considering the lesser-known Beltrami hemisphere model of hyperbolic geometry, we overcome the arithmetic limitations of Klein construction.
Finally, we characterize the bisectors and (pre)geodesics in the other Poincar\' e upper half-space, the Poincar\'e ball, and the Lorentz hyperboloid models, and discusses degenerate cases for which the dual hyperbolic Delaunay complex is not a triangulation.
\end{abstract}

{\bf Keywords}:
Hyperbolic geometry; 
Voronoi diagram; 
power diagram; 
affine diagram;
M\"obius diagram;
Delaunay complex;
Delaunay triangulation;
hemisphere model;
Klein ball model; 
Poincar\'e ball model; 
upper halfspace model; 
Lorentz hyperboloid model;
geodesic metric space.
\\

%%%%%%%%%
\section{Introduction}
%%%%%%%%%

Given a finite distinct point set $\calP=\{p_1,\ldots, p_n\}$ of a space $\bbX$ equipped with a distance function $d(\cdot,\cdot)$, the Voronoi diagram~\cite{compgeom-1998} of $\calP$ tessellates\footnote{Pairwise disjoint interior tiles fully covering $\bbX$.} the space into proximity regions called Voronoi cells.
The Voronoi cell of point $p$ is defined by

\begin{equation}
\vor(p):= \left\{  x\in\bbX \ \middle|\  d(p,x) \leq d(q,x),\ \forall q\in\calP\right\}, \forall p\in\calP.
\end{equation}

The ordinary $d$-dimensional Voronoi diagram is obtained by taking the Euclidean distance $d(p,q)=\|p-q\|_E:=\sqrt{\innerE{p-q}{p-q}}=\sqrt{\sum_{i=1}^d (p_i-q_i)^2}$ (square root of the Euclidean inner product of the vector difference).
This Euclidean Voronoi diagram has been extensively studied~\cite{okabeVD-1992} in relation with the Delaunay triangulation.
The Delaunay triangulation~\cite{Delaunay-1934} $\DT(\calP)$ of a point set $\calP$ is defined as a 
triangulation such that no point of $\calP$ falls strictly inside the circumscribing spheres of its simplices anchored at $\calP$. The Delaunay triangulation is unique for points in general position\footnote{That is,  no $k$-flat containing $k+2$ points of $\calP$ and no $k$-sphere passing through $k+3$ points of $\calP$ (for $1\leq k\leq d-1$).} (no collinear nor co-spherical degeneracies).

In this work, we consider the Voronoi diagram in the {\em $d$-dimensional real hyperbolic geometry} space $\bbX=\bbH^d(\kappa)$  of {\em constant negative sectional curvature}\footnote{For spherical geometry, the Gaussian curvature of the $d$-dimensional sphere of radius $r$ is $\kappa=\frac{1}{r^2}$. The curvature tends to $0$ (flat) when $r\rightarrow\infty$.} $\kappa$. We refer the reader to the survey~\cite{SurveyHyperbolicGeometry-1997} for a concise description of the five standard models of hyperbolic geometry with their relationships.
We call those five models as follows: 
\begin{itemize}
\item the Poincar\' e \underline{U}pper half-space (U), 
\item the \underline{P}oincar\' e ball (P), 
\item the \underline{K}lein ball (K), 
\item the \underline{L}orentz hyperboloid (L), and 
\item the \underline{B}eltrami hemisphere models (B).
\end{itemize} 
The above B/K/L/P/U naming and mnemonic of those models does not necessarily reflect the historical development of non-euclidean geometry.
See~\cite{HG-Milnor-1982} for a historical perspective of the first 150 years of hyperbolic geometry.
 
The paper is organized as follows:
Section~\ref{sec:prior} briefly recalls  prior work concerning the hyperbolic Voronoi diagram in arbitrary dimension, 
and section~\ref{sec:motivation} motivates the study of hyperbolic Voronoi diagrams by presenting  some applications ranging from computer graphics to  computational information geometry and machine learning.
Section~\ref{sec:hemisphere} describes the hyperbolic Voronoi diagram in the hemisphere model and show how to compute it using off-the-shelf algorithms relying only on rational arithmetic.
Section~\ref{sec:hvdmodels} describes the bisectors and (pre)geodesics for the five models, and discusses on degenerate cases for which the dual Delaunay complex it not a triangulation.
In \S\ref{sec:SSPD}, we present a connection between the spaces of symmetric positive-definite $2\times 2$ matrices of constant determinant and hyperbolic geometry. 
Finally, section~\ref{sec:conclusion} concludes this work.
Appendix~\ref{app:klein} reformulates our former work~\cite{KHVD-2010} in arbitrary dimension for sake of completeness.

%%%%%%%%%%%
\subsection{Prior work}\label{sec:prior}
%%%%%%%%%%%
Since we consider the general $d$-dimensional setting, let us recall only prior work tackling the hyperbolic Voronoi diagram constructions in {\em arbitrary dimension}.
Boissonnat and Yvinec~\cite{compgeom-1998} (pp. 449-454)
proved that the complexity of the hyperbolic Voronoi diagram of a $n$-point set $\calP\in\bbH^d(-1)$ is $\Theta(n^{\ceil{\frac{d}{2}}})$ using the Poincar\'e $d$-dimensional upper half-space model (U). 
They proceed by exhibiting an injective correspondence between the Euclidean and the hyperbolic diagrams using two successive projections.
Nielsen and Nock~\cite{KHVD-2010} showed that the hyperbolic Voronoi diagram in the Klein ball model (K) amounts to compute a clipped power diagram~\cite{powerdiagrams-1987}. Appendix~\ref{app:klein} recalls this construction, extending~\cite{KHVD-2010} to arbitrary dimension.
Bogdanov et al.~\cite{HDT-2011} proved that the hyperbolic Delaunay triangulation (dual graph\footnote{The Delaunay complex is however not always a triangulation when points are in degenerate positions. See Section~\ref{sec:hvdmodels}.} of the hyperbolic Voronoi diagram) in the Poincar\' e ball model (P) can be obtained by removing the simplices of the Euclidean Delaunay triangulation intersecting the bounding sphere~\cite{HDC-2020}.
Their approach requires only rational arithmetic.\footnote{However, we need to stick to the Poincar\'e model since conversion formula may introduce square root operations.}   

In this paper, we study the hyperbolic Voronoi diagram using the lesser known Beltrami hemisphere model~\cite{SurveyHyperbolicGeometry-1997} (B) and hyperboloid model (L), and show that the hyperbolic Voronoi diagram in those models amount to compute an affine diagram.

%%%%%%%%%%%%
\subsection{Applications}\label{sec:motivation}
%%%%%%%%%%%%
Hyperbolic Voronoi diagrams find applications in various fields of computer science.
In computer graphics, Walter~\cite{H-MDS-2004} proposed an hyperbolic image browser with its user-friendly interactive interface.
In network, Tanuma et al.~\cite{Tanuma-HVD-2011} extended the greedy geometric routing  of Kleinberg~\cite{HyperbolicRouting-2007} using the dual  structure of hyperbolic Voronoi diagrams.
In information geometry~\cite{KassVos-1997}, the Riemannian geometry induced by the Fisher information matrix of location-scale families of probability distributions (e.g., normal, Cauchy or Laplace distributions) amounts to hyperbolic geometry $\bbH^d(\kappa)$, where the  curvature constant $\kappa$ depends on the standard density of the location-scale family.
It follows that the statistical Voronoi diagrams with respect to the Fisher-Rao Riemannian distance~\cite{KassVos-1997} amounts to compute hyperbolic Voronoi diagrams.
Hyperbolic geometry has been used in machine learning for embedding hierarchical structures with low distortions (e.g.,~\cite{MLembedding-2018}).

%%%%%%%%%
\section{Hyperbolic Voronoi diagram in the Beltrami hemisphere model}\label{sec:hemisphere}
%%%%%%%%%

The $d$-dimensional hyperbolic space can be studied using various analytic models embedded in $\bbR^{d+1}$ using an extra dimension~\cite{SurveyHyperbolicGeometry-1997} denoted by $x_0$.
The hemisphere model $\calS^+\subset\bbR^{d+1}$ of hyperbolic space with constant curvature $\kappa=-\frac{1}{r^2}$ is defined on the positive  half-sphere of radius $r=\sqrt{-\frac{1}{\kappa}}$:

\begin{equation}
\calS^+ = \left\{ x\in\bbR^{d+1} \middle|\ \sum_{i=0}^{d} x_i^2=r^2=-\frac{1}{\kappa}\mbox{\ and\ } x_{0}>0\right\}.
\end{equation}

Although less prominent in the literature, this hemisphere model  was first reported by Beltrami~\cite{Beltrami-1868} in 1868. 
The hyperbolic distance between two points $p$ and $p'$ on  $\calS^+$ is expressed~\cite{HemisphereHyperbolicDistance-2001} in the hemisphere model by:

\begin{equation}\label{eq:dB}
d_B(p,p') = \arccosh \left( 1+\frac{1-\innerE{p}{p'}}{\innerE{p}{a}\innerE{p'}{a}} \right),
\end{equation}
where point $a$ denotes the southern pole of the sphere of radius $r=\sqrt{-\frac{1}{\kappa}}$: $a=(-r,0, ..., 0)$, $\arccosh(x)=\log(x+\sqrt{x^2-1})$ for $x\geq 1$, and $\innerE{p}{p'}=\sum_{i=0}^d p_ip_i'$ is the Euclidean inner product. 
We use the subscript $E$ to distinguish this inner product from the Lorentzian inner product introduced in Section~\ref{sec:hvdmodels}.

It follows from Eq.~\ref{eq:dB} that we have:

\begin{equation}
\cosh(d_B(p,p')) -1 = \frac{1-\inner{p}{p'}}{r^2 p_{0} p'_{0}},
\end{equation}
where $\cosh(x)=\frac{e^x+e^{-x}}{2}$ is the hyperbolic cosine function.
Thus the bisector $\Bi_B(p,p')$ of two points $p$ and $p'$ on the hemisphere model is  thus expressed by:

\begin{eqnarray}
\Bi_B(p,p') &:& \frac{1-\innerE{p}{x}}{r^2 p_{0}x_{0}} = \frac{1-\innerE{p'}{x}}{r^2 p'_{0}x_{0}},\\
%\Bi_B(p,p') &:& \InnerE{x}{\frac{p'}{p'_{0}} - \frac{p}{p_{0}}} + \frac{1}{p_{0}} - \frac{1}{p'_{0}} =0,\\
 \Bi_B(p,p') &:& \sum_{i=1}^{d} x_i \left(\frac{p'_i}{p'_{0}} - \frac{p_i}{p_{0}}\right)  + \frac{1}{p_{0}} - \frac{1}{p'_{0}} =0,\ x\in\calS^+.
\end{eqnarray}

Observe that  the bisector relaxed to $\bbR^{d+1}$ is a hyperplane with the coefficient corresponding to the term $x_{0}$ vanishing (that is, a vertical hyperplane). Therefore the bisector on the hemisphere is a vertical $(d-1)$-sphere.

To compute the hyperbolic Voronoi diagram on the Beltrami hemisphere, we can thus choose hyperplane $H_0: x_0=0$, compute the affine diagram, and clip it with the ball $\sum_{i=1}^d x_i^2<r^2$. We then either lift vertically this diagram onto the hemisphere, or manipulate the diagram internally using its affine representation on $H_0$. (For example, point location can be handled directly on $H_0$.) 

On hyperplane $H_0$, we write the bisector equation as:
\begin{equation}
\Bi_B(p,p') :  \left\langle x, \frac{p'}{p'_{0}} - \frac{p}{p_{0}} \right\rangle_{E^{d}}  + \frac{1}{p_{0}} - \frac{1}{p'_{0}} =0,
\end{equation}
where $\langle \cdot, \cdot \rangle_{E^{d}}$ denotes the $d$-dimensional Euclidean dot product on coordinates $x_1, ..., x_d$.
Affine diagrams can be built  equivalently using  power diagrams~\cite{powerdiagrams-1987}:
A power diagram   for a given a set $\calB=\{B_1, ..., B_n\}$ of $n$ balls (with $B_i=\ballE(c_i, w_i=r_i^2)$ for $i\in\{1, ..., n\}$), 
  is
defined as the minimization diagram~\cite{CurvedVoronoiDiagrams:2007} of the corresponding $n$ functions:
\begin{equation}
D_i(x)=\innerE{c_i-x}{c_i-x}-w_i,
\end{equation} 
where $w_i$ is the weight associated to center point $c_i$.
The power bisector $B_\pow(B_i,B_j)$ of
any two distinct balls $B_i=\ballE(c_i,w_i)$ and $B_j=\ballE(c_j,w_j)$ is the  {\em radical hyperplane}~\cite{powerdiagrams-1987} of equation:

\begin{equation}\label{eq:bipd}
\Bi_\pow(B_i,B_j): 2\innerE{x}{c_j-c_i}+ \innerE{c_i}{c_i}-  \innerE{c_j}{c_j}  + w_j - w_i=0.
\end{equation}
The last equation shows that power diagrams are additive weighted (ordinary) Voronoi diagrams~\cite{compgeom-1998}.

To find the equivalence of the Beltrami hemisphere affine diagram with a power diagram, we map the points $p$  to equivalent balls $B$ with centers $c$ and  weights $w$ such that:

$$
c=\frac{p}{2p_0},\quad w=\frac{1}{p_0}+\innerE{c}{c}=\frac{1}{p_0}+\frac{\innerE{p}{p}}{4p_0^2}.
$$ 
This calculation requires only rational arithmetic but input shall be given using $d+1$ coordinates.
(Klein hyperbolic Voronoi diagram used $d$-coordinates but requires algebraic coordinates, see Appendix~\ref{app:klein})

%Let us now identify the points on the upper hemisphere with half rays originating from the origin $o=(0, ..., 0)$.
%That is, points $p$ and $\lambda p$ belongs to the same half ray  for any $\lambda>0$.
%%This defines an equivalence class yielding projective geometry.
%Consider the homogeneous coordinates for $\tilde{p}=(p_1, ..., p_d | p_{0})$  and denote by
% $p^-=(\frac{p_{1}}{p_{0}}, ...,\frac{p_{d}}{p_{0}})$ its inhomogeneous coordinates.\footnote{For $p_0\not =0$, we can dehomogeneize $p$ in $\bbR^d$. 
% This   procedure  is called perspective division in computer vision. For the hemisphere model, we have $x_0>0$ by definition of $\calS^+$.}

Although the worst-case combinatorial complexity $O(n^{\ceil{\frac{d}{2}}})$ was already obtained in the Klein ball model~\cite{KHVD-2010}, the Poincar\'e ball model~\cite{HDT-2011}, and the Poincar\'e upper space model~\cite{compgeom-1998}, this novel hemisphere construction allows us to use off-the-shelf power diagram constructions to build the hyperbolic Voronoi diagram using only rational arithmetic.
Note that since the hyperbolic Voronoi diagram is a particular case of the Riemannian Voronoi diagram, it follows that the worst-case infinitesimally-scaled Euclidean Voronoi diagram (from equivalent polytopes on the $d+1$-dimensional curve of moments~\cite{compgeom-1998}) can be obtained. Therefore the complexity of the hyperbolic Voronoi diagram is $\Theta(n^{\ceil{\frac{d}{2}}})$.

% clipping and polar hyperplane

%There are several output-sensitive algorithms to build polytopes/convex hulls:
%
%\begin{itemize}
%
%\item Seidel~\cite{VorSeidel-1986} used a {\em shelling approach}  to get a $O(n^2+h\log n)$ running time.
%The quadratic start-up time was later reduced by Matousek and Schwarzkopf~\cite{LinearOptQueries-1993}.
%
%
%\item
%Using Chan's ray shooting algorithm~\cite{OSCH-1996} (gift wrapping method), the $h$ faces of the hyperbolic Voronoi diagram can be computed in time $O(n\log h+(nh)^{1-\frac{1}{\floor{\frac{d}{2}}+1}} \log^{O(1)}n )$. This is optimal~\cite{OSCH-1996} for $h=O(n^{{\floor{\frac{d}{2}}}^{-1} - \epsilon})$, for any $\epsilon>0$. In dimension $3$ and $4$, Chan et al.~\cite{Vor-CSY-1997} and Amato and Ramos~\cite{Vor-AmatoRamos-1996} obtained a $O(h\log^{d-1} n)$ time algorithm,  respectively.
%
%\item We can alternatively use a novel output-sensitive construction~\cite{VorOS-2012} that runs in $O(h\log n\log\Delta)$ (by simplifying a quality mesh~\cite{SparseVorRefinement-2006}), where $\Delta$ denotes the spread of the point set (ratio of the diameter over the closest distinct pair of points).
%
%\end{itemize}

%%%%%%%%%
\section{Voronoi bisectors and (pre)geodesics in the five standard models}\label{sec:hvdmodels} 
%%%%%%%%%
Hyperbolic geometry can be studied under the framework of geodesic metric spaces~\cite{MetricSpaceNegCurvature-2009}.
A metric space $(\bbX,d)$ is geodesic if and only if there exists a function $\gamma$ such that for any pair of points $(p,q)\in\bbX^2$ and any pair of scalar $(s,t)\in [0,1]^2$, we have:
$$
d(\gamma(p,q;s),\gamma(p,q;t))= |s-t|\ d(p,q),
$$
with $\gamma(p,q;0)=p$ and $\gamma(p,q;1)=q$.
Complete Riemannian manifolds are geodesic metric spaces.
Riemannian geodesics are parameterized curves with constant speed (or equivalently are parameterized by arc length).
Pregeodesics $\Gamma(p,q)$ are the traces of geodesics $\gamma(p,q;t)$: 
$$
\Gamma(p,q):=\{\gamma(p,q;t)\ :\ t\in [0,1]\}.
$$  

Table~\ref{tab:distance} summarizes the Riemannian metrics and the Riemannian distances for 
the Beltrami hemisphere (B), the Poincar\' e upper half-space (U), the Poincar\'e ball (P), the Klein ball (K),  and the Lorentz hyperboloid (L) models. 
The five model domains are ($r=\sqrt{-\frac{1}{\kappa}}$):

\begin{eqnarray}
\calU^+ &=& \left\{(x_0,r,x_2, ... x_d)\ \middle|\ x_0>0\right\} \mbox{\ (upper space)},\\
\calP &=& \left\{(0,x_1,..., x_d)\ \middle|\ \sum_{i=1}^d x_i^2<r^2\right\} \mbox{\ (Poincar\'e ball)},\\
\calS^+ &=& \left\{(x_0,x_1,..., x_d)\ \middle|\ \sum_{i=0}^d x_i^2<r^2\mbox{\ and\ } x_0>0 \right\} \mbox{\ (Beltrami hemisphere)},\\
\calK &=& \left\{(r,x_1,..., x_d)\ \middle|\ \sum_{i=1}^d x_i^2<r^2\right\}  \mbox{\ (Klein ball)},\\
\calL^+ &=& \left\{(x_0,x_1,..., x_d)\ \middle|\ \sum_{i=1}^d x_i^2-x_0^2=-r^2   \mbox{\ and\ } x_0>0\right\} \mbox{\ (Lorentz hyperboloid)}.
\end{eqnarray}

The Poincar\'e  and Klein ball models and the Poincar\'e upper half-space model have been often considered from an application point of view since  their domain dimension does not need to increase the ambient dimension and they can be easily displayed on a computer screen.

\begin{table}
\centering
\renewcommand{\arraystretch}{1.5}
\begin{tabular}{|l|l|c|l|}\hline
Model &  Riemannian metric $\ds^2$  & $\cosh \frac{d(p,q)}{\sqrt{-\kappa}}$, $d(\cdot,\cdot):$ Riemannian distance\\ \hline\hline
Klein (K)   &  $-\kappa \left( \frac{\ds_E^2}{1-\|x\|_E^2} + \frac{\innerE{x}{\dx}}{(1-\|x\|_E^2)^2} \right)$  & $   \frac{1-\innerE{p}{q}}{\sqrt{(1-\|p\|_E^2) (1-\|q\|_E^2)}}  $\\ \hline
Poincar\'e (P)   & $-\kappa \frac{4\ds_E^2}{(1-\|x\|_E^2)^2}$  & $   1+\frac{2\|p-q\|_E^2}{(1-\|p\|_E^2) (1-\|q\|_E^2)} $\\ \hline
Upper (U)  & $-\kappa\frac{\ds_E^2}{x_d^2}$  &   $1+ \frac{ \|p-q\|^2_E}{2p_d q_d}  $ \\ \hline
Hyperboloid (L) & $\kappa \dx_0^2+\ds_E^2$  & $\frac{\innerL{p}{q}}{\kappa}$  \\ \hline
Hemisphere (B) & $\frac{\sum_{i=0}^d \dx_i^2}{x_0^2}$  & $1+\frac{1-\innerE{p}{q}}{\kappa p_0q_0}$\\ \hline
\end{tabular}

\caption{Models of real hyperbolic geometry  $\bbH^d(\kappa)$ of constant section negative curvature $\kappa$ as Riemannian geometries: Riemannian metrics and distances.
All models are conformal (angle-preserving) except Klein model that is not conformal (except at the origin).
\label{tab:distance}
}
\end{table}

These five models are all linked in $\bbR^{d+1}$ as described in~\cite{SurveyHyperbolicGeometry-1997} (Figure~5, page 70). 
The maps from the hemisphere to the other models are either central or vertical projections explicited in~\cite{SurveyHyperbolicGeometry-1997}.
Let us summarize the bisector expressions in those five standard models:

\begin{eqnarray}
\Bi_B(p,q) &:&   \sum_{i=1}^{d} x_i \left(\frac{q_i}{q_{0}} - \frac{p_i}{p_{0}}\right)  + \frac{1}{p_{0}} - \frac{1}{q'_{0}} =0,\\
\Bi_K(p,q) &:& \InnerE{x}{q\sqrt{1-\|p\|_E^2}-p\sqrt{1-\|q\|_E^2}} + \sqrt{1-\|q\|_E^2} - \sqrt{1-\|p\|_E^2}=0,\\
\Bi_P(p,q) &:& \InnerE{x}{x} \left( \frac{1}{1-\|p\|_E^2} - \frac{1}{1-\|q\|_E^2} \right) +
2\InnerE{x}{\frac{q}{1-\|q\|^2_E} - \frac{p}{1-\|p\|^2_E}} +\nonumber\\
&& \frac{\|p\|_E^2}{1-\|p\|_E^2} - \frac{\|q\|_E^2}{1-\|q\|_E^2} =0,\\
\Bi_L(p,q) &:& \innerL{x}{q-p} =  (p_0-q_0) x_0 + \sum_{i=1}^d  (q_i-p_i) x_i =0,\\
\Bi_U(p,q) &:& \innerE{x}{x} \left(\frac{1}{p_d}-\frac{1}{q_d}\right) +2\InnerE{x}{\frac{q}{q_d}-\frac{p}{p_d}} + \frac{\innerE{p}{p}}{p_d}
-\frac{\innerE{q}{q}}{q_d}=0.\\
\end{eqnarray}
where $\innerL{x}{y}=-x_0y_0+\sum_{i=1}^d x_iy_i$ denotes the Lorentzian\footnote{Although attributed to Hendrik Lorentz, the model is due to Karl Weierstrass~\cite{DiacuCurvedEquilibria-2012}.} inner product.
Observe that the Lorentz bisectors relaxed to $\bbR^{d+1}$ are hyperplanes  all passing through the origin.
(The pregeodesics are hyperbola arcs contained in a hyperplane passing through the origin.)
The Lorentz hyperbolic Voronoi diagram can thus be obtained as the intersection of a $(d+1)$-dimensional affine diagram with the hyperboloid upper sheet $\bbL^+$. 
For odd dimension $d=2k+1$, we can thus build optimally the hyperbolic Voronoi diagram in the Lorentz model since 
$O(n^{\ceil{\frac{d+1}{2}}})=O(n^{\ceil{\frac{d}{2}}})=O(n^{k+1})$.
That is, the extra dimension $x_0$ for odd parity does not exhibit the dimension gap in the combinatorial complexity of the Voronoi diagram.
The hyperboloid model has been used by Galperin~\cite{Galperin-1993} to propose the {\em model centroid} in a closed-form expression.
(Indeed the hyperboloic K\"archer centroid~\cite{Karcher:1977} does not admit an analytic expression.)

Table~\ref{tab:vd} characterizes the nature of the bisectors (bisecting sites) and pregeodesics (linking sites) for those models.
 
\def\tabtext#1{\begin{minipage}{0.33\textwidth}#1\end{minipage}} 
\begin{table}
\centering

\begin{tabular}{|l||l|l|} \hline
Model & Pregeodesic & Bisector\\ \hline\hline
Hemisphere & arc of circles &  \tabtext{vertical spherical portion of the hemisphere}\\ \hline
Klein ball & line segment & hyperplane\\ \hline
Poincar\'e ball & \tabtext{  circular arc perpendicular to the bounding sphere, or straight lines passing through the origin}  & 
\tabtext{spherical portions, or hyperplanes passing through the origin when  $\|p\|=\|q\|$
}  \\ \hline
Upper half-space&
\tabtext{arc of circles or vertical lines}  & 
\tabtext{spherical portions or hyperplanes} \\ \hline
Hyperboloid & \tabtext{hyperbola} & \tabtext{intersection of hyperplane passing through the origin with the hyperboloid}\\ \hline
\end{tabular}

\caption{Hyperboloic Voronoi diagrams: Characterization of pregeodesics and bisectors for the five standard models.
\label{tab:vd}
}
\end{table}

The pregeodesics $\Gamma_K(p,q)$ in the Klein model are line segments expressed as linear interpolations of the endpoints:
$$
\Gamma_K(p,q)=\{(1-\beta)p+\beta q\ :\ \beta\in [0,1]\}.
$$

That is, the pregeodesics in the Klein model are obtained using the linear interpolation scheme $\LERP(p,q;\beta):=(1-\beta)p+\beta q$.
 That is, we have:
 $\Gamma_K(p,q)=\{\LERP(p,q,\beta)\ :\ \beta\in [0,1]\}$. 

Since the Klein model is a geodesic metric space $(\mathcal{K},d_K)$, let us report the expression of Klein geodesics so that we have
for any pair of points $(p,q)\in\mathcal{K}^2$ and any pair of scalar $(s,t)\in [0,1]^2$:
$$
d_K(\gamma_K(p,q;s),\gamma_K(p,q;t))= |s-t|\ d_K(p,q),
$$
with $\gamma_K(p,q;0)=p$ and $\gamma_K(p,q;1)=q$.

Recall that the Klein metric distance $d_K(p,q)$ between any two points $p$ and $q$ in the unit disk centered at the origin with curvature 
$\kappa=-1$ is
$$
d_K(p,q)=  \arccosh \left( \frac{1-p^\top q}{\sqrt{(1-p^\top p)}\sqrt{(1-q^\top q)}} \right).
$$

We have $d_K(p,\gamma_K(p,q;\alpha))=\alpha d_K(p,q)$ with $\gamma_K(p,q;\alpha)=(1-c(\alpha))p+c(\alpha) q$.
Thus we need to solve for $c(\alpha)$ in the equation:
$$
\frac{a-bc(\alpha)}{\sqrt{a(a-2bc(\alpha)+c'c(\alpha)^2)}}-d(\alpha)=0,
$$
with
\begin{eqnarray*}
a &:=& 1-p^\top p,\\
b &:=& p^\top (q-p),\\
c' &:=& (q-p)^\top (q-p) ,\\
d(\alpha) &:=& \cosh(\alpha d_K(p,q))
\end{eqnarray*}

Using symbolic calculations, we find the following solution:
 
\begin{equation}
c(\alpha) = \frac{   ad(\alpha) \sqrt{(ac'+b^2)  (d(\alpha)^2-1)} +a b (1-d(\alpha)^2)}{a c' d(\alpha)^2 + b^2}.
\end{equation}

Thus we get in closed-form the Klein geodesics $\gamma_K(p,q;\alpha)$ (albeit a large formula).

We check that our solution yields for all $(p,q)\in\mathcal{K}^2$ and $(s,t)\in [0,1]^2$:
$$
d_K(\gamma_K(p,q;s),\gamma_K(p,q;t))= |s-t|\ d_K(p,q),\quad \forall s,t\in [0,1].
$$

We can use the Klein geodesics to efficiently compute the smallest enclosing ball of a finite point set $\calP$ in hyperbolic geometry~\cite{ArnaudonNielsen-2012}. With the expression of the Klein geodesic, we do not need to perform hyperbolic translations from and to the disk origin as in~\cite{NielsenHadjeres-2015}.
The algorithm for calculating an approximation of the smallest enclosing ball of $\calP=\{p_1,\ldots, p_n\}$ is:
\begin{itemize}
\item Initialize $c_1=p_1$
\item Repeat $t$ times: Let $c_{i+1}=\gamma_K\left(c_i,p_{f_i},\frac{1}{i+1}\right)$ where $p_{f_i}$ is the farthest point of $\calP$ to $c_i$.
That is, we have $f_i=\arg\max_{j\in\{1,\ldots, n\}} d_K(c_i,p_j)$.
\end{itemize}

The algorithm is proven to converge in~\cite{ArnaudonNielsen-2012} since the hyperbolic geometry is a Hadamard space. 

\begin{figure}
\centering

Figure~\ref{fig:degenerate} depicts some degenerate cases in the ball models for co-spherical points centered at the origin, and their corresponding diagrams on the upper half-space model.
 
\begin{tabular}{lcccc}
&Klein ball & Poincar\'e ball & Upper halfspace\\
(a) &\includegraphics[bb=0 0 1024 1024, width=0.25\textwidth]{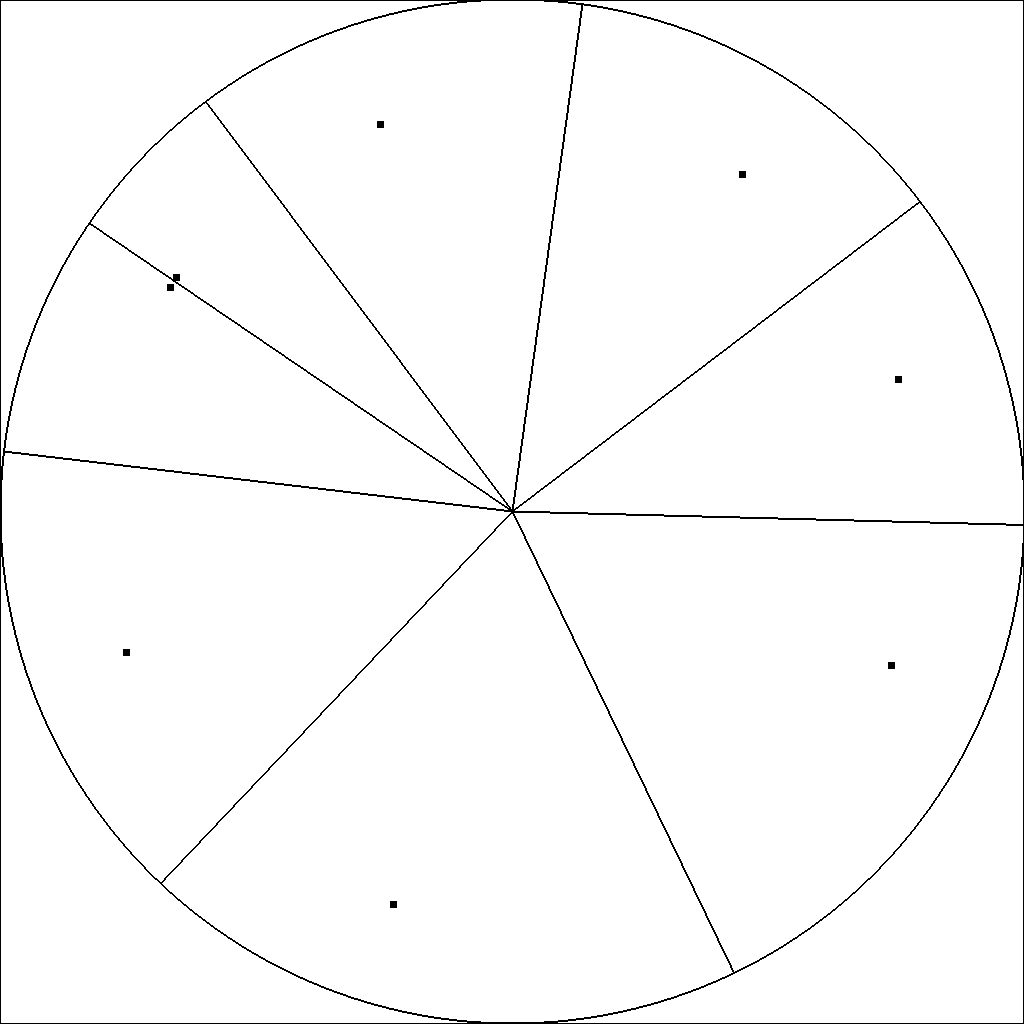} &
\includegraphics[bb=0 0 1024 1024, width=0.25\textwidth]{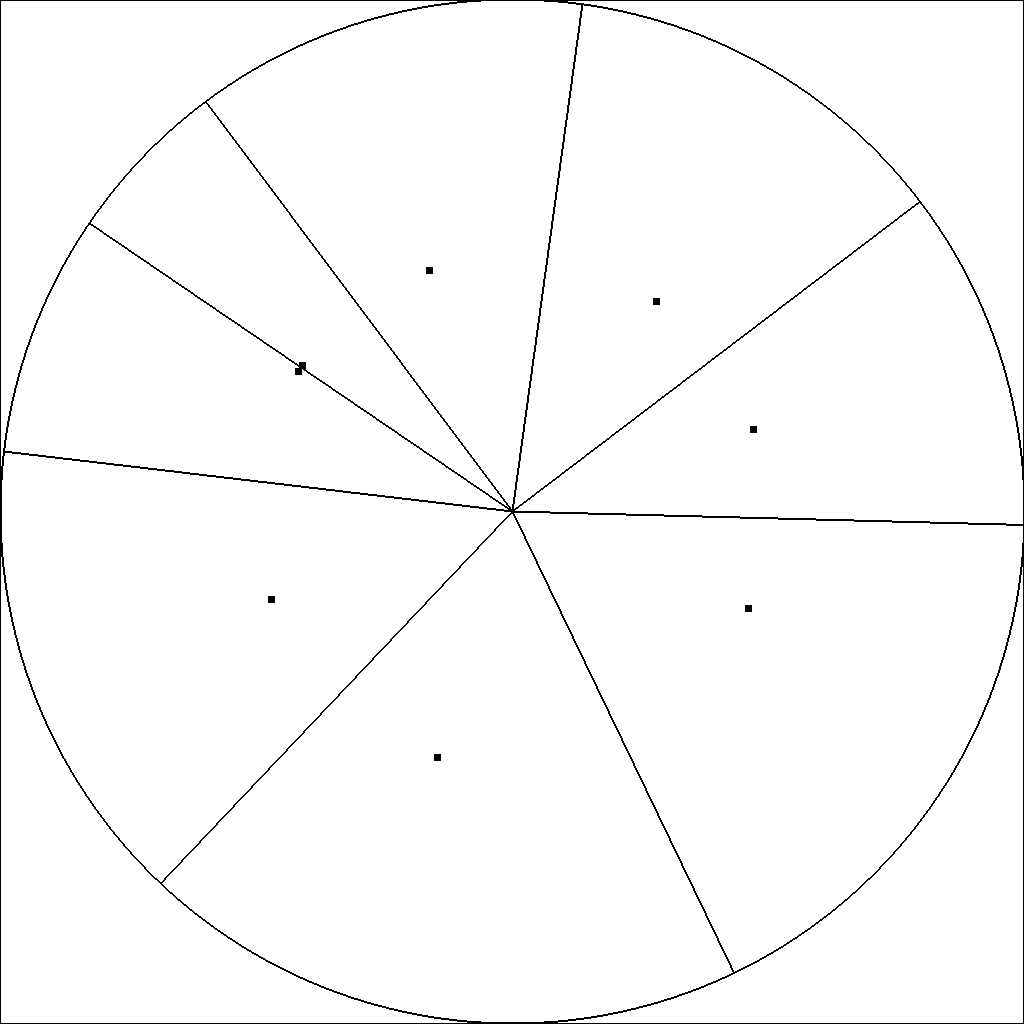} &
\includegraphics[bb=0 0 1024 1024, width=0.25\textwidth]{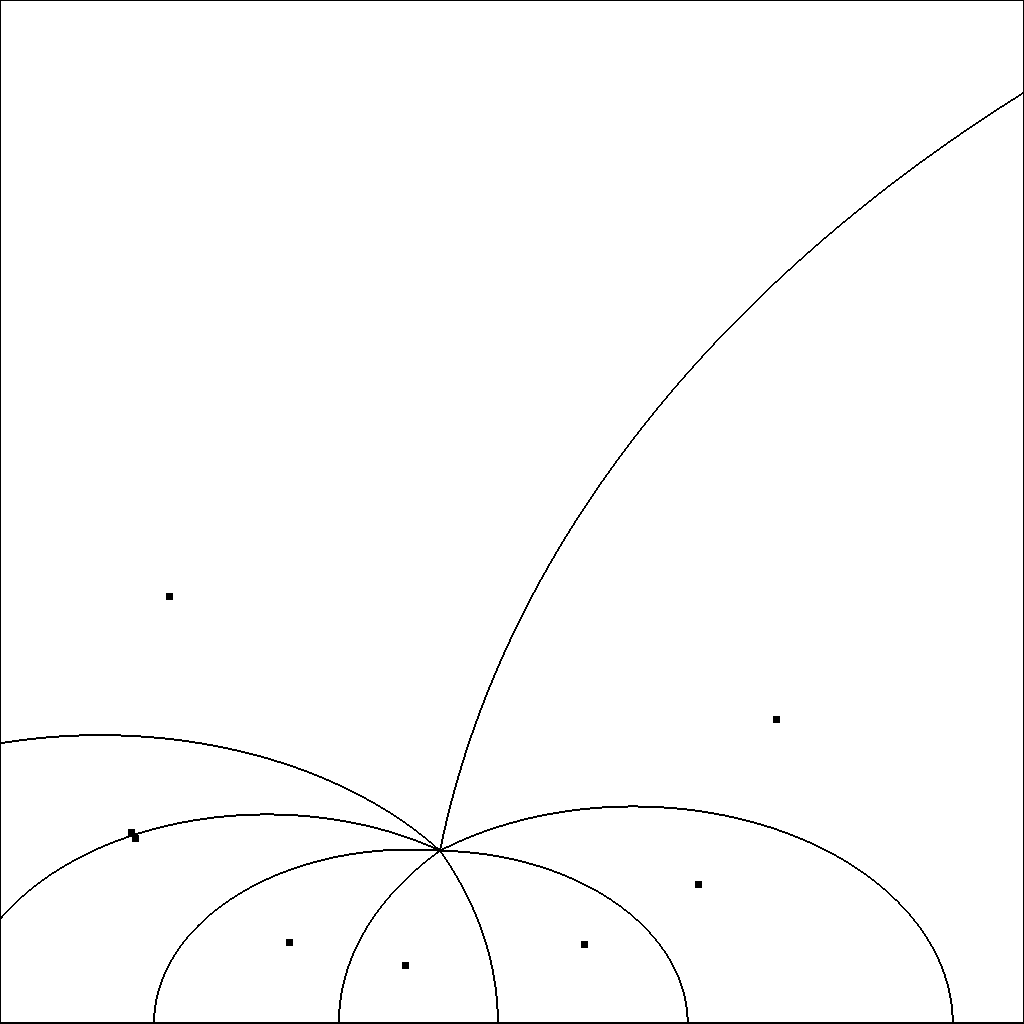}\\
(b)&\includegraphics[bb=0 0 1024 1024, width=0.25\textwidth]{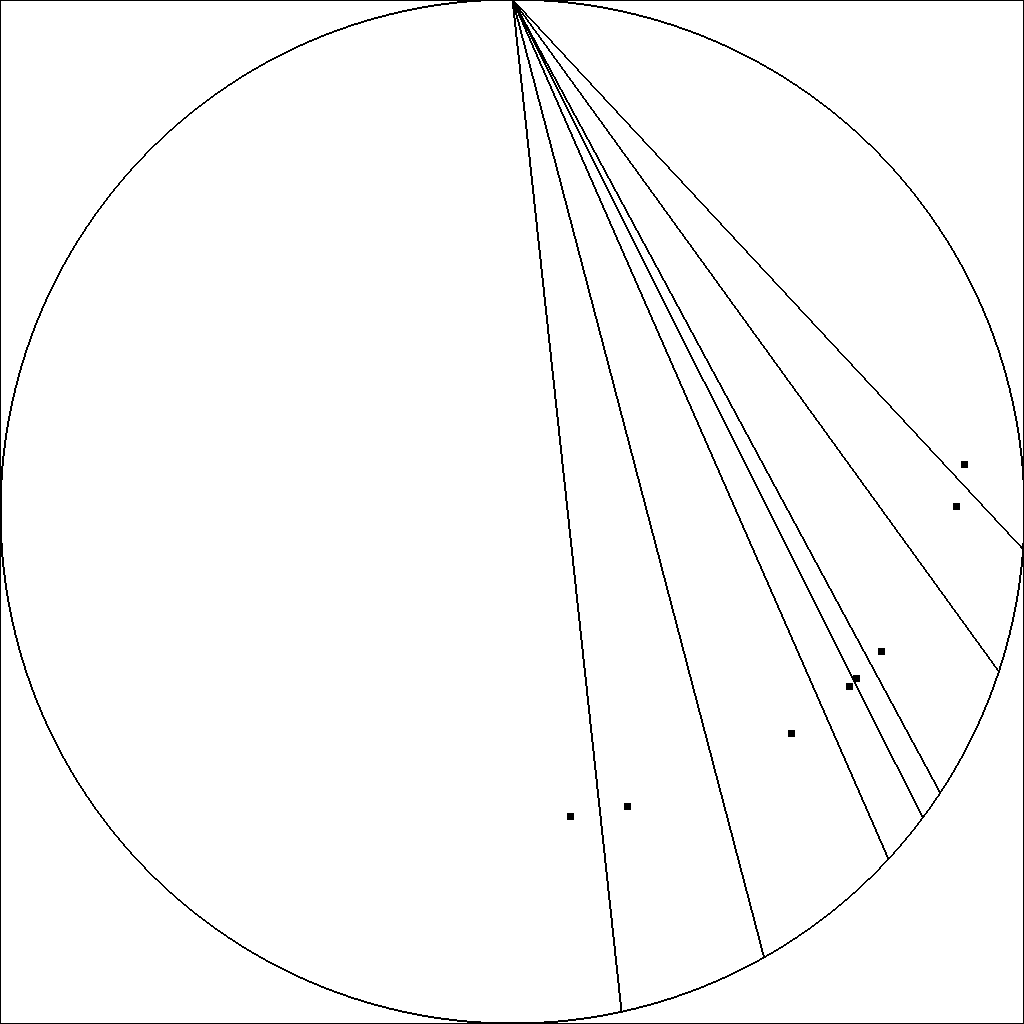} &
\includegraphics[bb=0 0 1024 1024, width=0.25\textwidth]{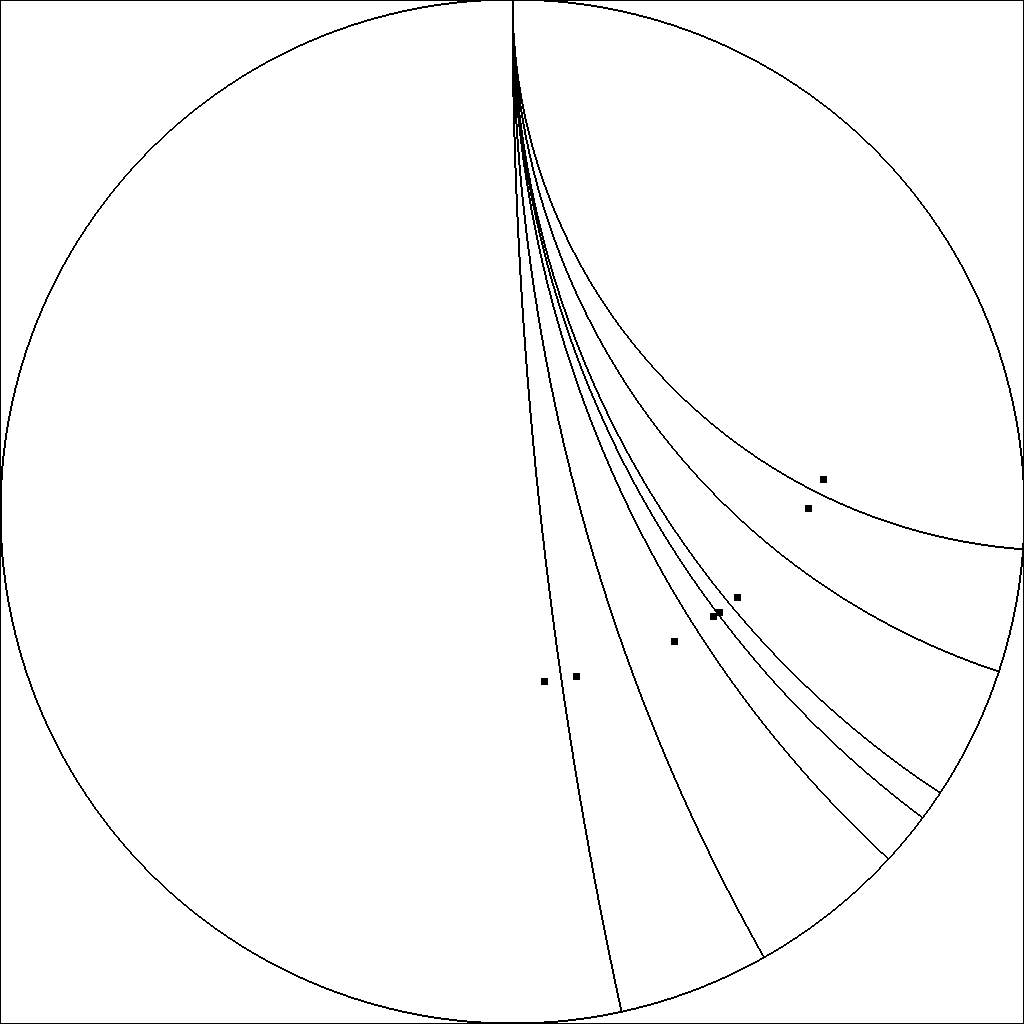} &
\includegraphics[bb=0 0 1024 1024, width=0.25\textwidth]{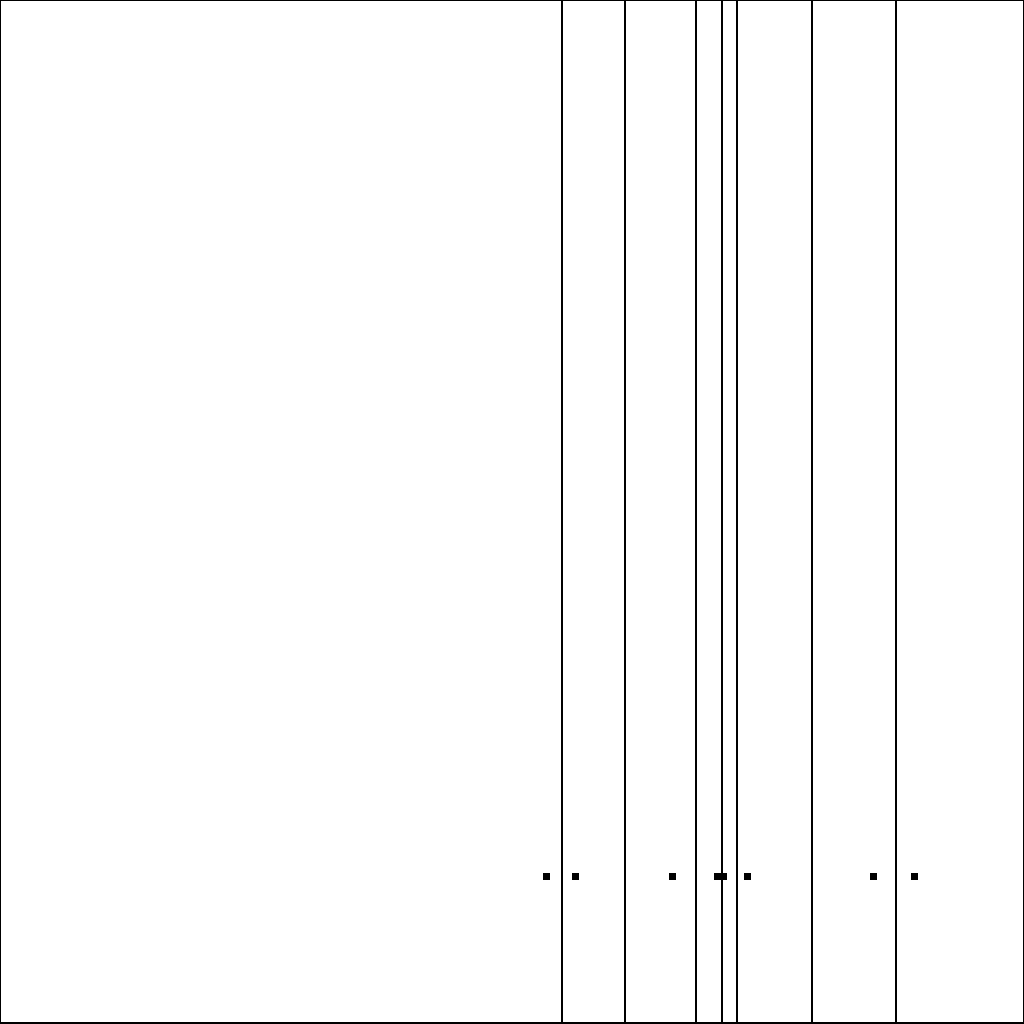}\\
\end{tabular}

\caption{Degenerate hyperbolic Voronoi diagrams:
(a): In the Poincar\'e/Klein ball models when point radii are all equal (wheel diagram with bisectors passing through the origin.)
(b): In the upper halfspace when the $d$-th coordinate are all equal.
\label{fig:degenerate}}
\end{figure} 

\begin{figure}
\centering
\begin{tabular}{cccc}
(a) \includegraphics[bb=0 0 1024 1024, width=0.3\textwidth]{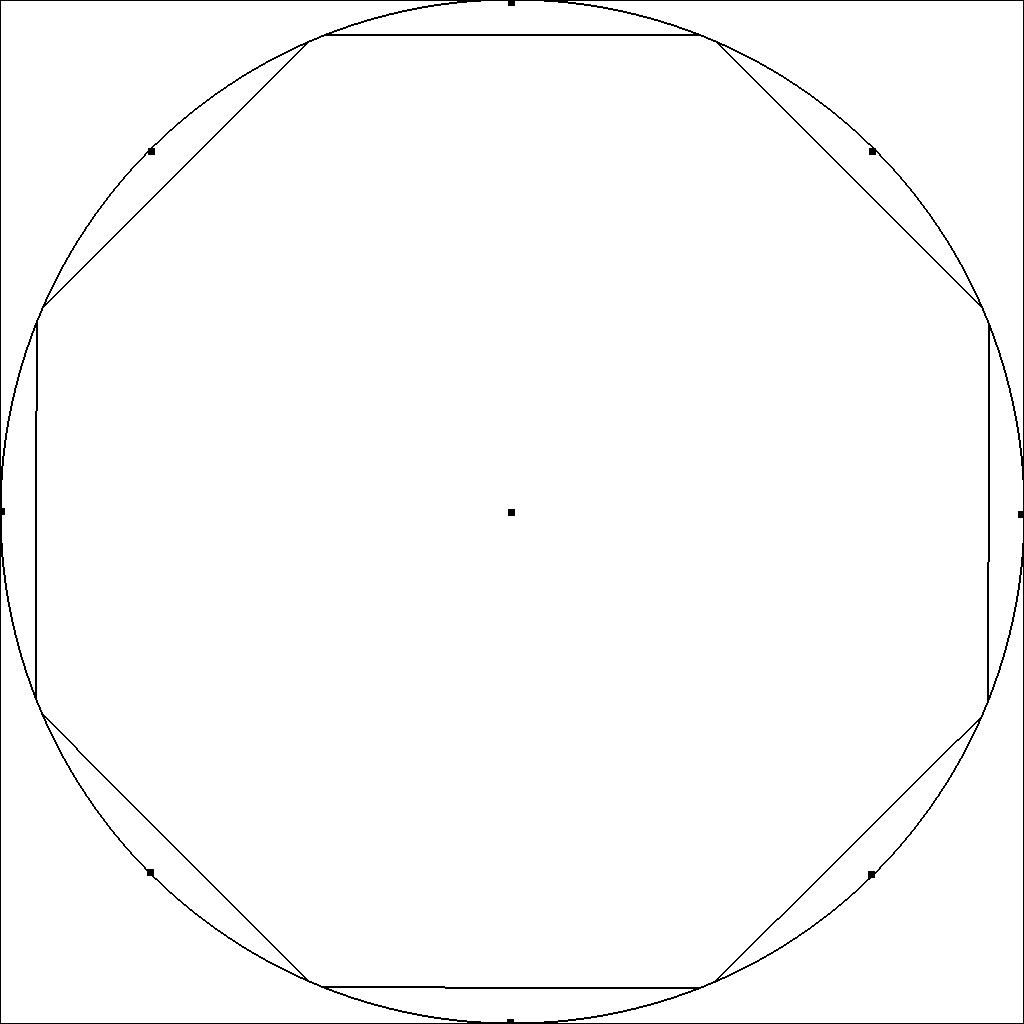}&
(b) \includegraphics[bb=0 0 1024 1024, width=0.3\textwidth]{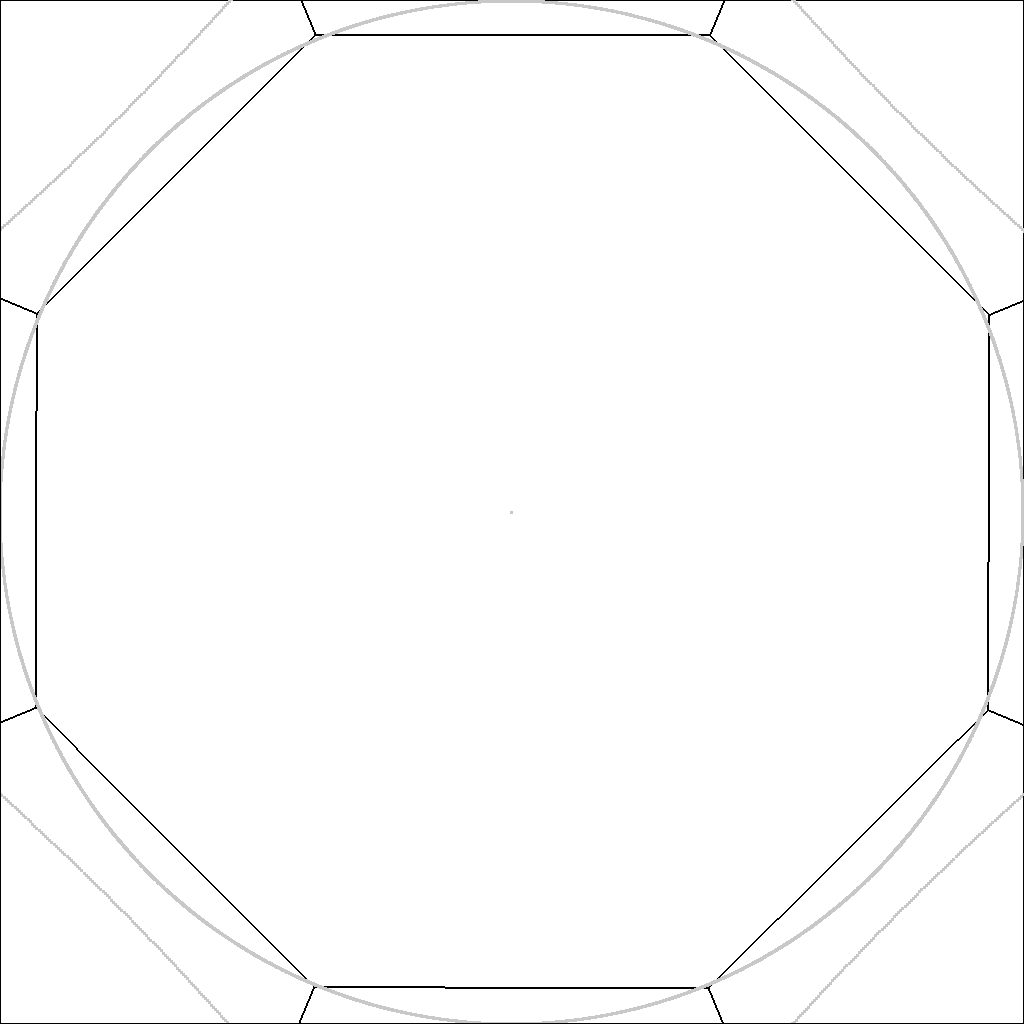} \\
(c) \includegraphics[bb=0 0 1024 1024, width=0.3\textwidth]{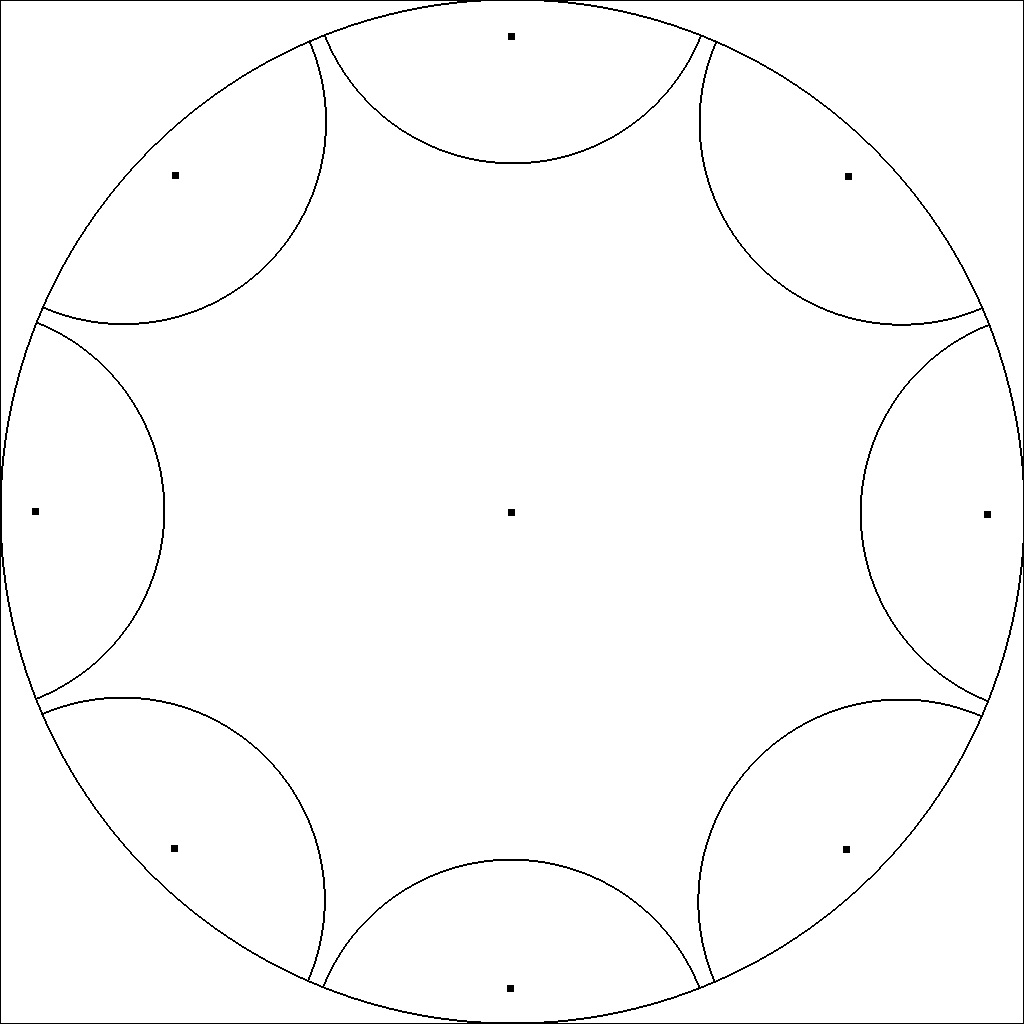}& 
(d) \includegraphics[bb=0 0 1024 1024, width=0.3\textwidth]{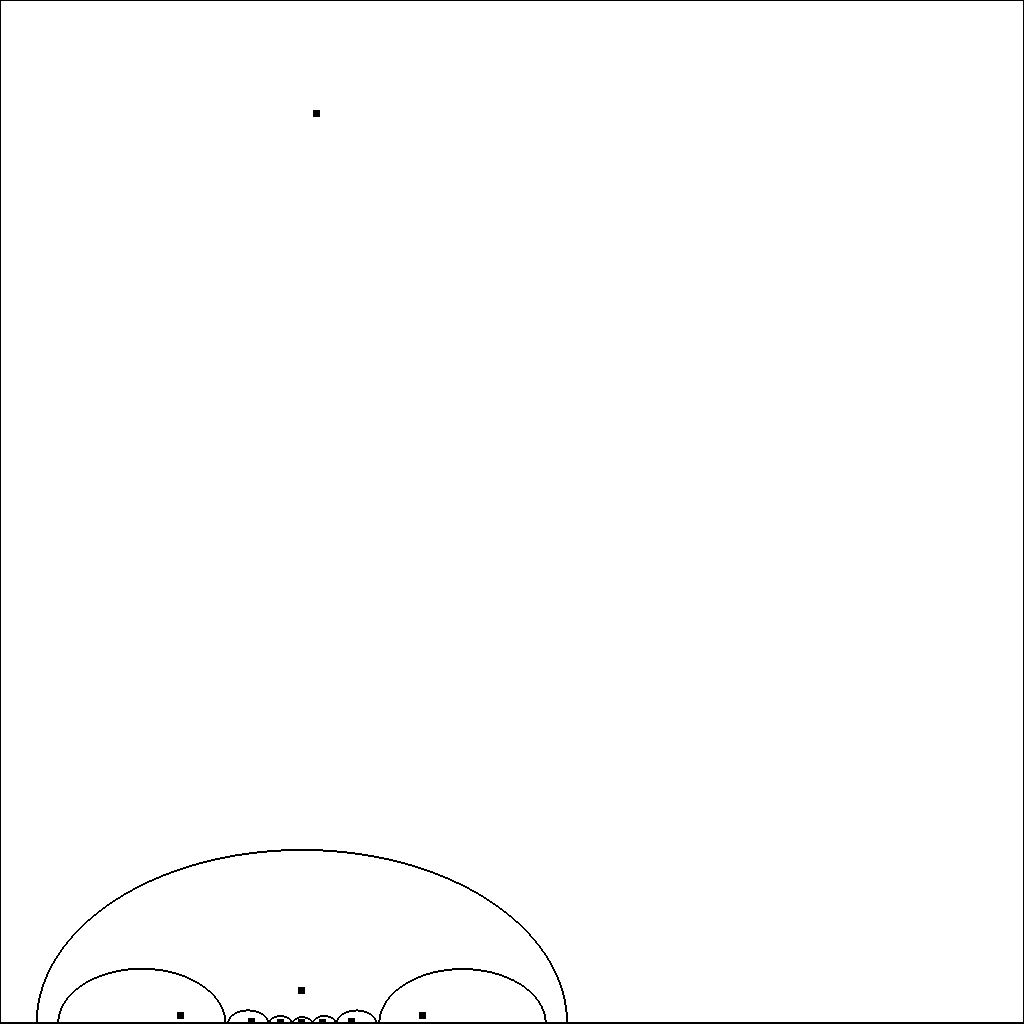}\\
\end{tabular}

\caption{A hyperbolic Voronoi diagram with all cells unbounded ($n=9$).
The dual graph corresponds to a tree:
(a) Klein, (b) equivalent power diagram, (c) Poincar\'e ball, and (d) Poincar\'e upper halfspace.
\label{fig:unbounded}
}

\end{figure}

It is well-known that for points in general position (no collinear nor co-spherical points), the dual of an ordinary Voronoi diagram is the unique Delaunay triangulation~\cite{compgeom-1998}. This duality holds for the hyperbolic Voronoi diagram and the hyperbolic Delaunay triangulation (with the empty sphere property), provided general position of the sites.
Hyperbolic Voronoi diagrams can be fairly different from ordinary Voronoi diagram.
For example, Figure~\ref{fig:unbounded} gives an example of the hyperbolic Voronoi diagram when all but one site are co-circular and close to the bounding circle with a single site centered at the ball center.
In that case the dual Delaunay complex is not a triangulation although unique (a $(n-1)$-ary tree of depth $1$). 
We refer to~\cite{StabilityDelaunay-2012} for a stability analysis of the Delaunay triangulation on generic manifolds.

\def\SPD{\mathrm{SPD}}
\def\SSPD{\mathrm{SSPD}}
\def\arccosh{\mathrm{arccosh}}
\def\arctanh{\mathrm{arctanh}}
\def\innerlabel#1#2#3{{\langle #1,#2\rangle_{#3}}}
\def\Log{\mathrm{Log}}
\def\dP{\mathrm{d}P}
\def\tr{\mathrm{tr}}
\def\mattwotwo#1#2#3#4{\left[\begin{array}{ll}#1 & #2 \cr #3 & #4\end{array}\right]}
%%%%%%%%%
\section{Hyperbolic geometry of the special symmetric positive-definite $2\times 2$ matrices (SSPD)}\label{sec:SSPD}
%%%%%%%%%

Let $\SPD=\{P\in \mathrm{Mat}_\bbR(d,d)\ :\ P\succ 0\}$ denote the set of symmetric positive-definite (SPD) $d\times d$ matrices.
Consider $\SPD$ as a Riemannian manifold~\cite{SPD-2011} $(M,g)$ equipped with the trace metric $g$ defining at each tangent plane $T_P$  
the following inner product:
$$
\forall A,B\in T_P,\quad g(A,B)=\innerlabel{A}{B}{P}=\tr(P^{-1}AP^{-1}B).
$$
The SPD manifold is an open convex cone with geodesic equation $\ddot{P}-\dot{P}P^{-1}\dot{P}=0$, and
he geodesic $P_{12}(t)$ passing through $P_1$ and $P_2$ is parameterized as follows:
$$
P_{12}(t)=P_1^{\frac{1}{2}} \exp(t \Log(P_1^{-\frac{1}{2}} P_2 P_1^{-\frac{1}{2}})) P_1^{\frac{1}{2}},
$$
where $\Log(P)$ denotes the unique matrix logarithm of $P$   since all eigenvalues of $P$ are positive.
The Riemannian induced distance is
$$
\rho_\SPD(P_1,P_2)=\sqrt{\sum_{i=1}^d \log^2 \lambda_i(P_1P_2^{-1})},
$$
where $\lambda_i(P)>0$ denotes the $i$-th largest eigenvalue of matrix of $P\in\SPD$.
Observe that the Riemannian trace metric distance is scale-invariant:
$$
\forall \lambda>0,\quad \rho_\SPD(\lambda P_1,\lambda P_2)=\rho_\SPD(P_1,P_2).
$$

When $d=2$, we have
$$
\SPD_2 = \left\{
 P=P(a,b,c)=\mattwotwo{a}{c}{c}{b}, a>0, ab-c^2>0
\right\},
$$
and the largest eigenvalue $\lambda_1$ and smallest eigenvalue $\lambda_2$ can be calculated as follows:
\begin{eqnarray*}
\lambda_1(P) &=&\frac{1}{2}\left(\tr(P)+\sqrt{\tr^2(P)-4|P|}\right)\\
\lambda_2(P) &=&\frac{1}{2}\left(\tr(P)-\sqrt{\tr^2(P)-4|P|}\right),
\end{eqnarray*}
where 
$\tr(P)=a+b$ denotes the matrix trace and  $|P|=ab-c^2$ the matrix determinant.

The space $\SPD_2$ can be foliated~\cite{Foliation-2006} (De Rham decomposition) as follows:
$$
\SPD_2=\SSPD_2(r)\times\bbR_{++},
$$
where
$$
\SSPD_2(r) = \left\{
\mattwotwo{a}{c}{c}{b}, a>0, ab-c^2=r
\right\}.
$$

The space $\SSPD_2(r)$ denotes the space of special symmetric positive-definite matrices (SSPD) with prescribed determinant equal to $r$.
For any $r>0$, the $\SSPD_2(r)$  is a totally geodesic submanifold of constant negative curvature $\kappa=-1/2$, and thus is isometric to the 2D hyperbolic geometry with curvature $-1/2$.
It follows that we have for matrices $P_1$ and $P_2$ in $\SSPD(r)$ the following expression of the Riemannian matrix trace distance:
\begin{equation}
\rho_{\SSPD_2}(P_1,P_2)=
\rho_{\SSPD_2}\left( \frac{P_1}{\sqrt{|P_1|}},\frac{P_2}{\sqrt{|P_2|}}\right)=
\sqrt{2}\,\rho_K\left(k\left(\frac{P_1}{\sqrt{|P_1|}}\right),k\left(\frac{P_2}{\sqrt{|P_2|}}\right)\right),
\end{equation}
where $\rho_K$ denotes the hyperbolic distance with curvature $-1$ expressed using the Klein unit disk model:
$$
\rho_K(k_1,k_2)=\arccosh\left(\frac{1-k_1\cdot k_2}{\sqrt{1-k_1\cdot k_1}\sqrt{1-k_2\cdot k_2}}\right),
$$
where $p\cdot q=\innerlabel{p}{q}{E}$ denotes the Euclidean inner product (dot product).
The map $k:\SPD\rightarrow \bbR^2$ transforms positive-definite matrices $P=P(a,b,c)$ to corresponding points on the Klein unit disk as follows:
\begin{equation}
k(P)=\left(\frac{a^2+2a-b^2-2b}{a^2+2a+b^2+2b+2+2c^2},\frac{2c(a+b+2)}{a^2+2a+b^2+2b+2+2c^2}\right).
\end{equation}
Here, notice that we considered the Klein unit disk for the hyperbolic geometry model of curvature $\kappa<0$, and we have:
$$
\rho_{K,\kappa}(k_1,k_2)=\sqrt{-\frac{1}{\kappa}}\,\rho_K(k_1,k_2).
$$
Thus, we have $\rho_{K,-\frac{1}{2}}(k_1,k_2)=\sqrt{2}\, \arccosh\left(\frac{1-k_1\cdot k_2}{\sqrt{1-k_1\cdot k_1}\sqrt{1-k_2\cdot k_2}}\right)$.

Instead of using the Klein model, we can also use the Poincar\'e disk (P) model
\begin{equation}
\forall P_1,P_2\in \SSPD_2(r),\quad \rho_{\SSPD_2}(P_1,P_2)=\sqrt{2}\,\rho_P\left(w\left(\frac{P_1}{\sqrt{|P_1|}}\right),w\left(\frac{P_2}{\sqrt{|P_2|}}\right)\right),
\end{equation}
where
$$
\rho_{P}(z_1,z_2) = 2\,\arctanh\left(\frac{|z_1-z_2|}{|1-\bar{z}_1z_2|}\right),
$$
$\arctanh(x)=\frac{1}{2}\log \frac{1+x}{1-x}$, and $w:\SPD_2\rightarrow\mathbb{C}$ denotes the mapping: 
\begin{equation}
w(P)= \frac{a-b}{2+a+b}+i\frac{c}{2+a+b}.
\end{equation}

We can also use the hyperboloid model (L), and have
\begin{equation}
\forall P_1,P_2\in \SSPD_2(r),\quad \rho_{\SSPD_2}(P_1,P_2)=\sqrt{2}\,\rho_L\left(h\left(\frac{P_1}{\sqrt{|P_1|}}\right),h\left(\frac{P_2}{\sqrt{|P_2|}}\right)\right),
\end{equation}
where
$$
\rho_L(h_1,h_2)=\arccosh\left(-\innerlabel{h_1}{h_2}{L}\right),
$$
with $\innerlabel{p}{q}{L}=-p_1q_1+p_2q_2+p_3q_3$, and $h:\SPD_2\rightarrow\bbR^3$ denotes the mapping of SPD matrices onto the upper sheet of the hyperboloid:
\begin{equation}
h(P)=\left[\begin{array}{c}
\frac{a+b}{2}\\
\frac{a-b}{2}\\
c
\end{array}
\right]
\end{equation}

When $d>2$, the space $\SPD_d$ can be foliated~\cite{Foliation-2006} as $\SPD_d=\SSPD_d\times \bbR_{++}$ (De Rham decomposition).
SPD is interpreted as a foliated manifold with leaves $\SSPD_d(r)$ of codimension $1$,
 where $\SSPD_d(r)$ denotes the space of  special symmetric positive-definite $d\times d$ matrices with prescribed determinant equal to $r$.
We can interpret $\SSPD_d$ as the following quotient space:
$$
\SSPD_d(1) \cong \mathrm{SL}(d)/\mathrm{SO}(d).
$$
The spaces $\SSPD_d(r)$ are Einstein submanifolds of $\SPD_d$.

%%%%%%%%%
\section{Conclusion}\label{sec:conclusion}
%%%%%%%%%
In the Klein ball model, we extended  the planar case~\cite{KHVD-2010} and proved that the hyperbolic Voronoi diagram of a finite point set in arbitrary dimension is affine. 
It followed a simple method to optimally calculate the Klein hyperbolic Voronoi diagram by computing a corresponding power diagram clipped to the bounding sphere.
This method however required non-rational arithmetic (square root operations in the denominator) on $d$-dimensional coordinates as noticed in~\cite{HDT-2011}.
To overcome this drawback, we described a novel approach in the hemisphere model, and showed that the hyperbolic Voronoi diagram also amounts to compute an affine diagram using only rational arithmetic on $(d+1)$-dimensional coordinates. 
From the view point of computational geometry, those various analytical models of hyperbolic space allow one to take advantage of the merits of each model. 
The Poincar\'e/Klein ball models are both conformal\footnote{Preserving angles. That is, the metric is locally a scaled Euclidean metric (identity matrix).} at the origin with geodesics passing through the origin being straight lines.
For geodesic walking-type algorithms like computing the hyperbolic centroid, Barbaresco~\cite{Barbaresco-2008} considered the Poincar\'e ball model and perform a hyperbolic translation at each step to set the current point to the origin in order to ensure that geodesics departing from that current point   are straight line segments. This could have been also performed using the Klein model.
When manipulating these models, we may perform a hyperbolic rigid motion~\cite{HemisphereHyperbolicDistance-2001} to choose a convenient origin~\cite{MoebiusTransformation-2001}.\footnote{Orientation preserving isometries in 2D consists of all fractional linear transformations with real coefficients and unit determinant. That is the group of M\"obius transformations  ($\PSL_2\bbR$). The real hyperbolic plane is then identified with the complex hyperbolic line.
Orientation preserving isometries in 3D can be identified with $\PSL_2\bbC$.
That is, all fractional linear transformations of the plane of points at infinity with complex coefficients, see~\cite{HG-Milnor-1982}.
}

%%%%%%%%%
\section*{Acknowledgments}
%%%%%%%%%
Frank Nielsen would like to acknowledge email correspondence with Dr. Olivier Devillers on this topic~\cite{HDT-2012} that considers the space of spheres~\cite{BVD-2010} for tackling the hyperbolic Voronoi and hyperbolic Delaunay complex~\cite{HDC-2020}.

%\bibliographystyle{plain}
%\bibliography{HyperbolicGeometry}

\appendix

\section{Hyperbolic Voronoi diagrams in the Klein projective ball model}\label{app:klein}

This appendix extends~\cite{KHVD-2010} to arbitrary dimension. 
In the Klein ball model $\calK(\kappa)$, the hyperbolic geometry of curvature $\kappa=-\frac{1}{r^2}$ is embedded~\cite{MetricSpaceNegCurvature-2009} inside a Euclidean ball 
$\bbB=\{x\in\mathbb{R}^d\ |\ \|x\|_E<r \}$ of radius $r$ with the corresponding Klein hyperbolic distance  $d_K(p,q)$ between two points $p$ and $q$ expressed by:

\begin{equation}
d_K(p,q) = \sqrt{-\kappa}\times \arccosh \left(\frac{1-\innerE{p}{q}}{\sqrt{(1-\|p\|_E^2) (1-\|q\|_E^2)}}  \right),
\end{equation}
where $\arccosh(x)=\log(x+\sqrt{x^2-1})$ for $x\geq 1$. 

Consider the Klein bisector $\Bi_K(p,q)$ of any two distinct points $p$ and $q$:

\begin{equation}
\Bi_K(p,q) = \left\{ x\in \calK(\kappa) \middle|\ d_K(p,x) = d_K(q,x) \right\},
\end{equation}

Since $\arccosh(x)$  is a monotonic function preserving the distance order and $\sqrt{-\kappa}$ is a multiplicative constant, it follows that:

\begin{equation}
\Bi_K(p,q)           = \left\{ x\in\bbB^d  \middle| 
           \frac{1-\innerE{p}{x}}{\sqrt{(1-\|p\|_E^2) (1-\|x\|_E^2)}}  = 
           \frac{1-\innerE{q}{x}}{\sqrt{(1-\|q\|_E^2) (1-\|x\|_E)^2}} \right\}.
\end{equation}

Therefore, the bisector $B_K(p,q)$ is the loci of the points $x$ satisfying:

\begin{equation}\label{eq:kb} 
\InnerE{x}{q\sqrt{1-\|p\|_E^2}-p\sqrt{1-\|q\|_E^2}} + \sqrt{1-\|q\|_E^2} - \sqrt{1-\|p\|_E^2}=0.
\end{equation}

This is a linear equation $\innerE{a}{x}+b=0$ in $x$, with:   
\begin{equation}
a = q\sqrt{1-\|p\|_E^2}-p\sqrt{1-\|q\|_E^2},\quad
b = \sqrt{1-\|q\|_E^2}-\sqrt{1-\|p\|_E^2}.
\end{equation}

That is,  Klein bisectors are  hyperplanes.
It follows that the Klein hyperbolic Voronoi diagram is an {\em affine diagram}~\cite{CurvedVoronoiDiagrams:2007} with all its Voronoi cells convex.\footnote{It is a particular affine diagram since all its cells are guaranteed non-empty.}
Affine diagrams can be {\em universally} built from an equivalent power diagram~\cite{CurvedVoronoiDiagrams:2007}.
The power distance $d_\pow(B,x)$ of a point $x$ to a Euclidean ball
$B=\ballE(c,r)$ of center $c$   and radius $r$ is defined as:

\begin{equation}
d_\pow(B,x) = \innerE{c-x}{c-x}-r^2.\label{eq:pd}
\end{equation} 

Given a set $\calB=\{B_1, ..., B_n\}$ of $n$ balls with $B_i=\ballE(c_i, r_i)$, 
 the power diagram  is
defined as the minimization diagram~\cite{CurvedVoronoiDiagrams:2007} of the corresponding $n$ functions:
\begin{equation}
D_i(x)=\innerE{c_i-x}{c_i-x}-r_i^2.
\end{equation} 
The power bisector $B_\pow(B_i,B_j)$ of
any two distinct balls $B_i=\ballE(c_i,r_i)$ and $B_j=\ballE(c_j,r_j)$ is the  {\em radical hyperplane}~\cite{powerdiagrams-1987} of equation:

\begin{equation}
\Bi_\pow(B_i,B_j): 2\innerE{x}{c_j-c_i}+ \innerE{c_i}{c_i}-  \innerE{c_j}{c_j}  + r_j^2-r_i^2=0.
\end{equation}

\begin{figure}
\centering
\begin{tabular}{cccc}
(a) \includegraphics[bb=0 0 1024 1024, width=0.3\textwidth]{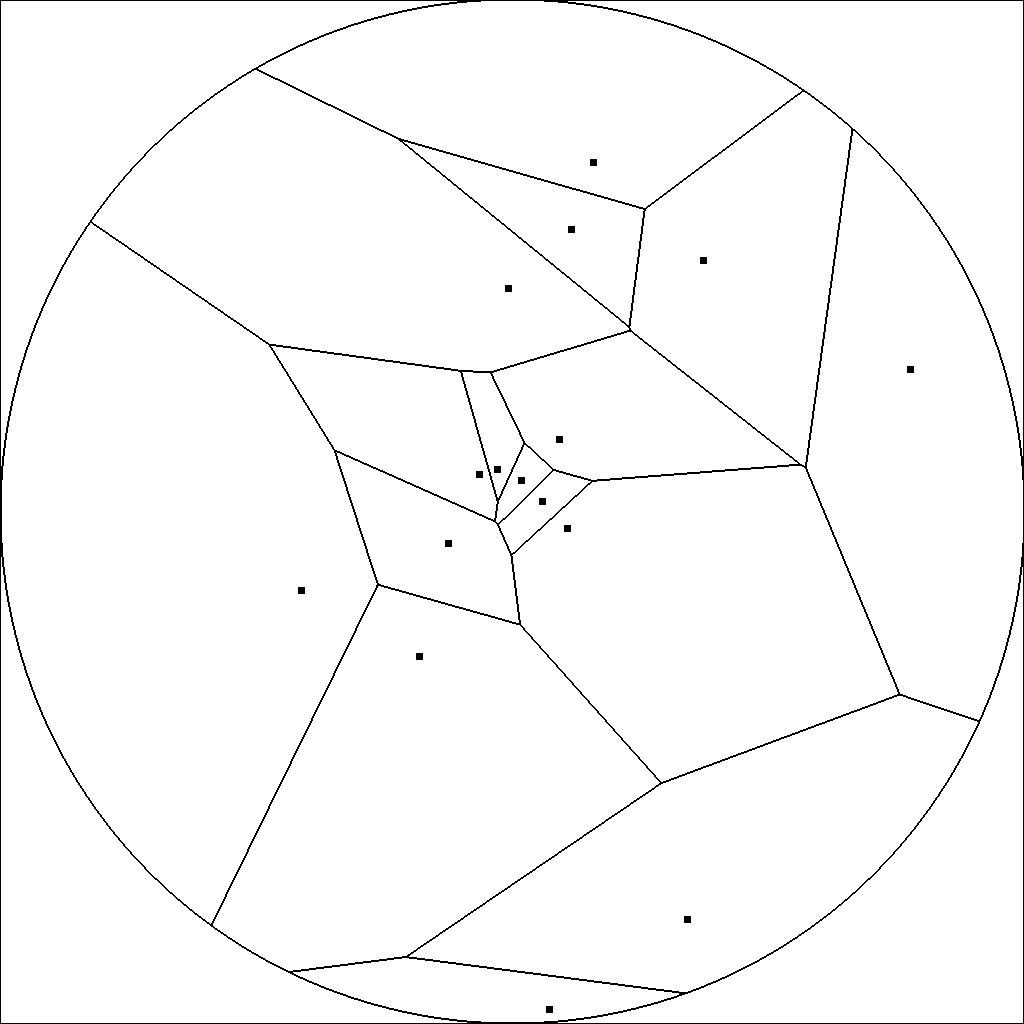}&
(b) \includegraphics[bb=0 0 1024 1024, width=0.3\textwidth]{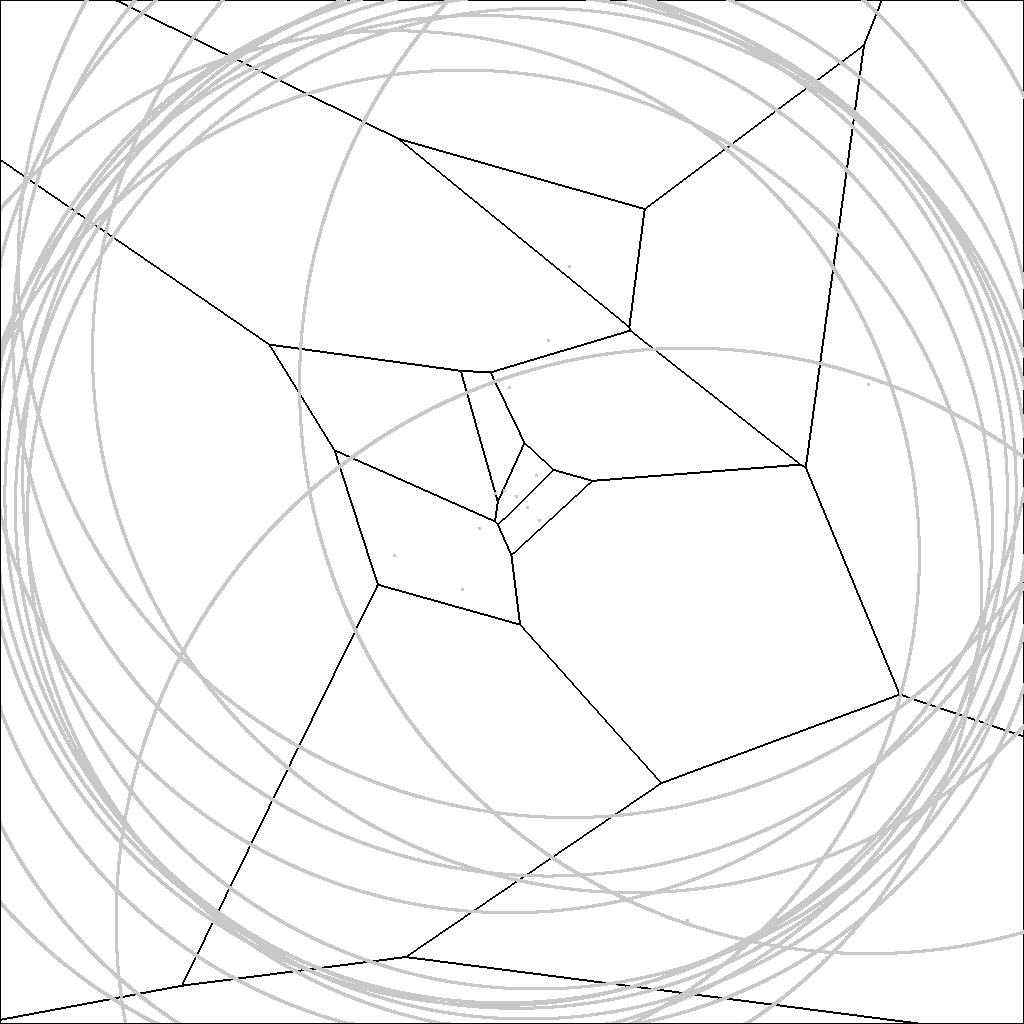} \\
(c) \includegraphics[bb=0 0 1024 1024, width=0.3\textwidth]{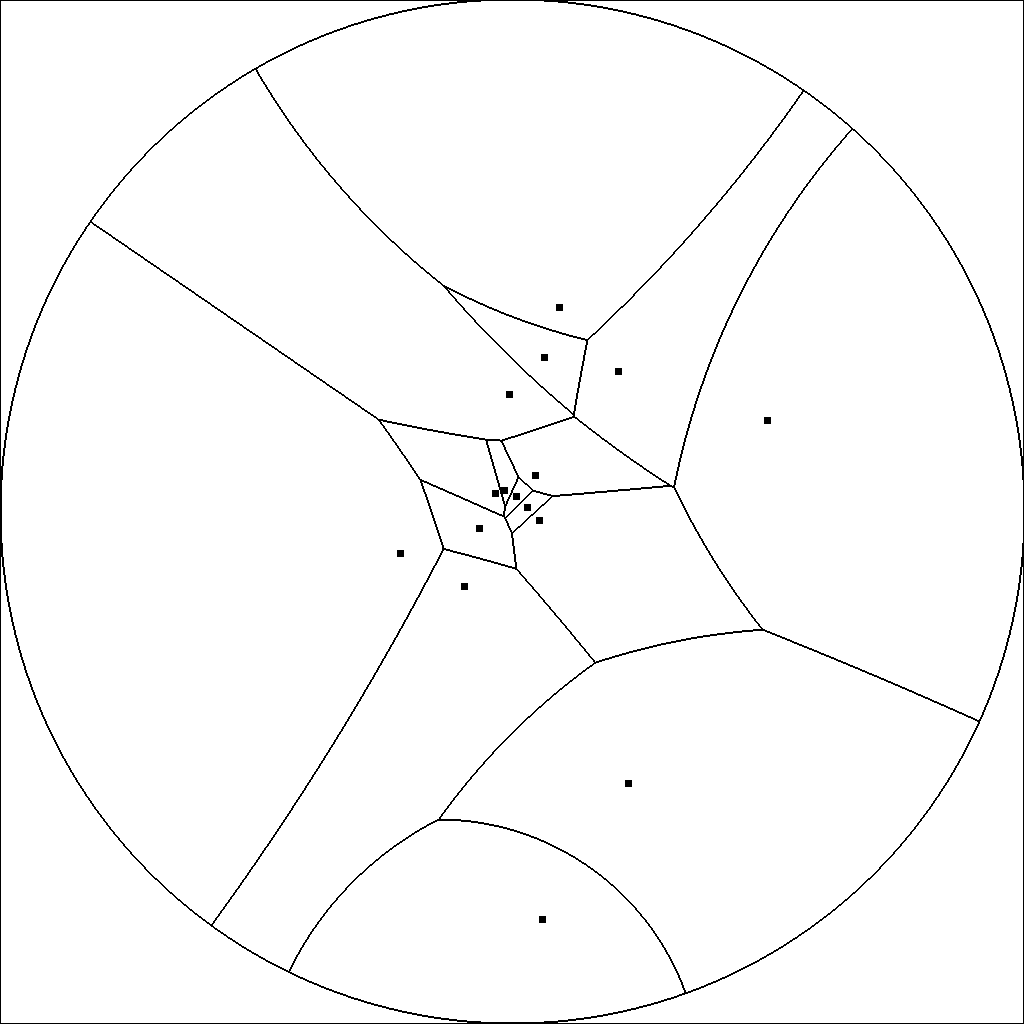}& 
(d) \includegraphics[bb=0 0 1024 1024, width=0.3\textwidth]{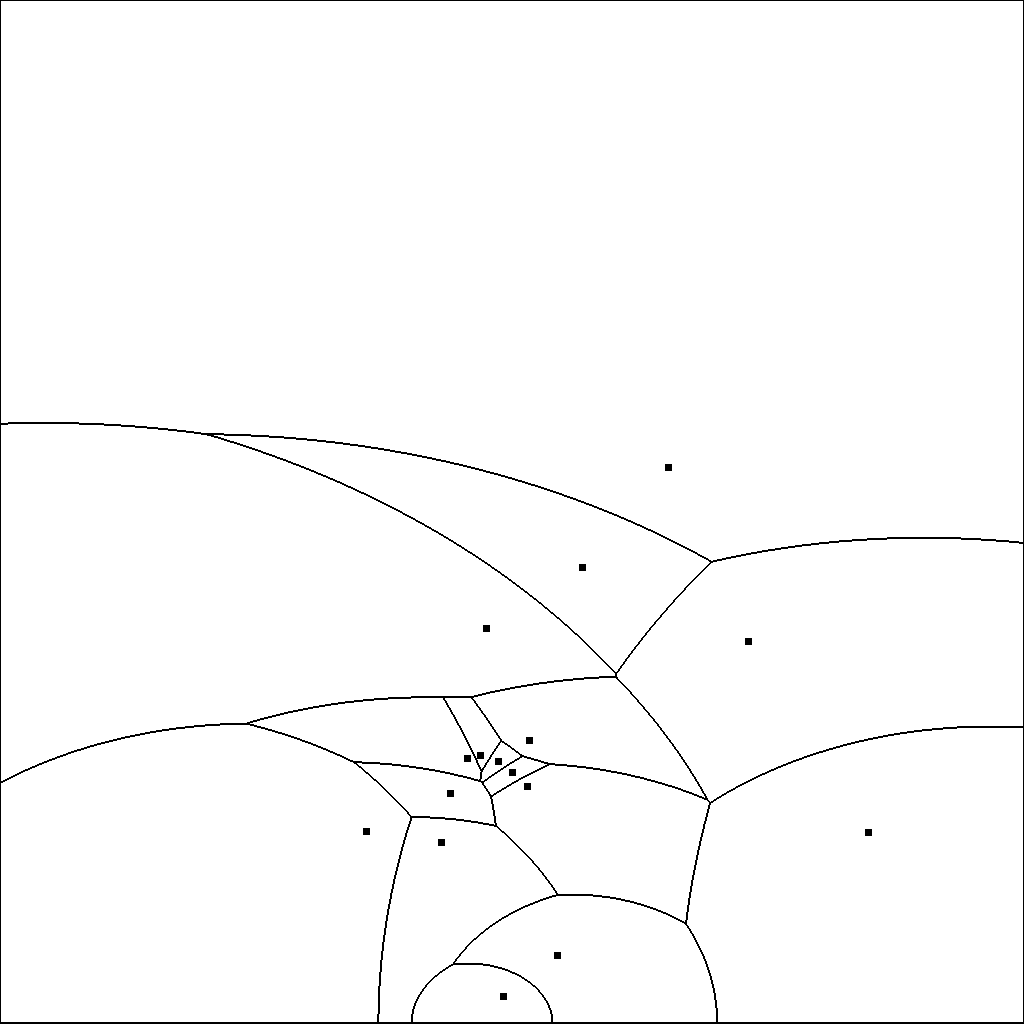}\\
\end{tabular}
\caption{(a) The Klein Voronoi diagram calculated from the restriction of the power diagram (b) to the  unit ball.
Converting the Klein diagram into the Poincar\'e Voronoi diagram (c) and the upper half-space Voronoi diagram.
 Observe that the combinatorial complexity of the cells does not change ($n=16$ points).
  The power diagram may have  cells empty of sites and ball centers may be located outside the square of side length $r=\sqrt{-\frac{1}{\kappa}}$.  
\label{fig:vorklein}}
\end{figure}

This power bisector linear equation shows that power diagrams are also affine diagrams.
Let us report the mapping that associates to points in the Klein ball  corresponding Euclidean balls so that their bisectors coincide pairwise. 
By identifying the bisector equation Eq.~\ref{eq:kb} with that of Eq.~\ref{eq:bipd}, we find that  point $p_i$ is mapped to a ball $B_i$ of center $c_i$ with squared radius $r_i^2$, where:

\begin{equation}
c_i=c(B_i)=\frac{p_i}{2\sqrt{1-\|p_i\|_E^2}},\quad
r_i^2=r^2(B_i)=\frac{\|p_i\|_E^2}{4(1-\|p_i\|_E^2)}-\frac{1}{\sqrt{1-\|p_i\|_E^2}}.\label{eq:radius}
\end{equation}
Observe that the ball centers may span the full $d$-dimensional Euclidean space $\bbR^d$.
The radii may be negative and are interpreted as balls with imaginary radii (or as weighted points with real weights, see Eq.~\ref{eq:pd}).
Solving for the quadratic expression\footnote{Let $l=\|p\|_E^2\in [0,1]$. We seek when $\frac{l}{4(1-l)}-\frac{1}{\sqrt{1-l}}<0$. That is,
 we solve for the root of equation $l-4\sqrt{1-l}$ in $[0,1]$. This is a quadratic expression $l^2+16l-16=0$ which admits a unique root in $[0,1]$: $4(\sqrt{5}-2)$.} of Eq.~\ref{eq:radius}, we find that $r^2(p_i)<0$  if and only if: 

\begin{equation}
\|p_i\|_E^2 < 4(\sqrt{5}-2)\simeq 0.944272.
\end{equation}
%(In fact, the squared of radii are better interpreted as additive weights $w_i=r_i^2$ on the squared Euclidean distance.)
%Since the squared Euclidean distance is a Bregman divergence, the power diagram can also be interpreted as an additive Bregman Voronoi diagram~\cite{bvd-2010} that are affine.
Thus to compute the Klein hyperbolic Voronoi diagram of point set $\calP=\{p_1, ..., p_n\}$, we first construct a set of corresponding balls  
\begin{equation}
\calB=\{B_1=\ballE(c(p_1),r(p_1), ..., B_n=\ballE(c(p_n),r(p_n)\},
\end{equation}
 compute
the power diagram\footnote{Power diagrams can be computed in arbitrary dimension  as convex hulls or halfspace intersections (unbounded polytope) using free software like  {\sf QHull} (\url{http://qhull.org}), {\sf CGAL} (\url{http://www.cgal.org}) or LEDA (\url{http://www.algorithmic-solutions.com/leda/}). } of $\calB$, and consider the restriction of this diagram to the unit  ball $\bbB^d$.
Figure~\ref{fig:vorklein}(a) and Figure~\ref{fig:vorklein}(b)  depict the Klein Voronoi diagram and corresponding power diagram, respectively.
It follows another proof of the combinatorial complexity of the hyperbolic Voronoi diagram using the Klein model instead of the Poincar\'e upper halfspace~\cite{compgeom-1998} (pp. 449-454) or the Poincar\'e ball model~\cite{HDT-2011}:

\begin{theorem}
The hyperbolic Voronoi diagram of $n$ $d$-dimensional points can be obtained  in the Klein ball model as an equivalent power diagram clipped with the interior of the unit ball.
The diagram has combinatorial complexity $O(n^{\ceil{\frac{d}{2}}})$ and can be built  in optimal
$O(n\log n+n^\ceil{\frac{d}{2}})$ time. 
\end{theorem}

Figure~\ref{fig:vorklein}(a)  displays the Klein hyperbolic Voronoi diagram built from an equivalent power diagram (Figure~\ref{fig:vorklein}(b)).
Note that all bounded Voronoi cells necessarily fall inside the ball, and therefore the construction can be made output-sensitive using Chan's algorithm~\cite{OSCH-1996}.

The following section describes the mapping functions to convert from the Klein model et the Poincar\'e ball, upper halfspace and Lorentz models.
  
%%%%%%%%%  
\section{Converting Klein model to other models}\label{sec:convertK}
%%%%%%%%%  
We report formula\footnote{Wlog., we assume $\kappa=-1$. Otherwise, we first rescale by $\sqrt{-\frac{1}{\kappa}}$.} and their inverse for  converting from/to Klein coordinates   into the Poincar\'e ball (P), upper halfspace (U) and Lorentz coordinates(L):

\begin{eqnarray}
P_K(x)&=&\frac{x}{1+\sqrt{1-\|x\|_E^2}},\\
K_P(x)&=&\frac{2x}{1+\|x\|_E^2},\\
U_K(x)&=&\frac{\left( 1-\frac{\|x\|_E^2}{1+\sqrt{1-\|x\|_E^2}}, x_1, ..., x_{d-1} \right)}{1-x_d},\\
K_U(x)&=&\frac{\left(\|x\|_E^2-1,2x_1, ..., 2x_{d-1}\right)}{1+\|x\|_E^2},\\
L_K(x)&=&\frac{(1,x_1, ..., x_d)}{\sqrt{1-\|x\|_E^2}},\\ \label{eq:LK}
K_L(x)&=&\left(\frac{x_1}{x_0}, ..., \frac{x_d}{x_0}\right). \label{eq:KL}
\end{eqnarray}  

\end{document}